\documentclass{elsart}
\usepackage{amssymb}
\usepackage{amsfonts}
\usepackage{amsmath}
\usepackage{graphicx}

\journal{Physics Letters A}

\begin{document}

\begin{frontmatter}

\title{Continuous-Variable Quantum Games \thanksref{publish}}

\thanks[publish]{{\em Physics Letters A} {\bf 306}, 73-78 (2002).}

\author[mphy]{Hui Li},
\ead{lhuy@mail.ustc.edu.cn}
\author[mphy,sgp]{Jiangfeng Du}, and
\ead{djf@ustc.edu.cn}
\author[bel]{Serge Massar}

\address[mphy]{Department of Modern Physics, University of Science and Technology of China,
Hefei, 230027, People's Republic China}
\address[bel]{Service de Physique Th\'{e}orique CP225, Universit\'{e} Libre de Bruxelles,
1050 Brussels, Belgium}
\address[sgp]{Department of Physics, National University of Singapore, 10 Kent Ridge Crescent, 119260, Singapore}

\begin{abstract}
We investigate the quantization of games in which the players can access to
a continuous set of classical strategies, making use of continuous-variable
quantum systems. For the particular case of the Cournot's Duopoly, we find
that, even though the two players both act as \textquotedblleft
selfishly\textquotedblright\ in the quantum game\ as they do in the
classical game, they are found to virtually cooperate due to the quantum
entanglement between them. We also find that the original
Einstein-Podolksy-Rosen state contributes to the best profits that the two
firms could ever attain. Moreover, we propose a practical experimental setup
for the implementation of such quantum games.
\end{abstract}


\end{frontmatter}

\section*{Introduction}

Quantum game theory was initiated by a paper of D. A. Meyer\cite{1},
discovering that a player can always beats his classical opponent by
adopting quantum strategies. Consequently the quantization of the famous
Prisoners' Dilemma was presented by J. Eisert \textit{et al.}\cite{2}. They
showed that this game ceases to pose a dilemma if quantum strategies are
allowed for. By far, investigations on multiplayer quantum games\cite{3,3-1}%
, as well as many interesting aspects on quantum games\cite%
{4,4-1,4-2,4-3,4-4,4-5,6}, were also presented. Recently, we investigated
the role of entanglement in quantum games and found that the properties of
the quantum Prisoners' Dilemma change discontinuously when its entanglement
varies\cite{7,8}. Besides the theoretical investigations, the first
experimental realization of quantum games was also successfully accomplished
on our NMR quantum computer\cite{8}.

However most of the current investigations on quantum games focus on games
in which the players have finite number of strategies. In real-life society
and economy, many cases should be represented by games in which the players
can access to a continuous set of strategies\cite{14}. Thus games with a
continuous set of strategies are quite familiar and are very important.
Considering the intimate connection between the theory of games and the
theory of quantum communication\cite{2}, quantization of games with
continuum strategic space therefore deserves thorough investigations, for
both theoretical and practical reasons.

In this paper we investigate the quantization of games with continuous
strategic space. A classic instance of such games is the Cournot's Duopoly%
\cite{9}, which is also a cornerstone of modern economics. The classical
game exhibits a dilemma-like situation, \textit{i.e.} the unique Nash
equilibrium is inferior to the \textit{Pareto Optimal}, just like in the
Prisoners' Dilemma\cite{2}. We give a quantum structure of Cournot's
Duopoly, and show that --- in the context of this structure --- the players
can escape the frustrating dilemma-like situation if the structure involves
a maximally entangled state. We also observe the transition of the game from
purely classical to fully quantum, as the game's entanglement increases from
zero to maximum. The profits at quantum Nash equilibrium increases
monotonously as the game's entanglement increases. In the maximally
entangled game, the profits at the unique Nash equilibrium is exactly the
Pareto Optimal (the best result they can ever attain while maintaining the
symmetry of the game), and the dilemma-like situation in the classical
Cournot's Duopoly is completely resolved. Moreover we propose a practical
experimental setup for the quantum Cournot's Duopoly, within the capability
of nowadays optical technology.

\section{Classical Cournot's Duopoly}

A duopoly is a case in which two firm monopolize the market of a certain
commodity without a third competitor. In a simple scenario of Cournot's
model for this duopoly, firm $1$ and firm $2$ simultaneously decide the
quantities $q_{1}$ and $q_{2}$, respectively, of a homogeneous product they
want to put on the market. Let $Q$ be the total quantity, \textit{i.e.} $%
Q=q_{1}+q_{2}$, and $P\left( Q\right) $ the price. A formula of $P\left(
Q\right) $ is $P\left( Q\right) =a-Q$ for $Q\leqslant a$ while $P\left(
Q\right) =0$ for $Q>a$. This is a simple but reasonable reflection of the
fact that the more product is put on the market, the less the price will be,
and when the total quantity is extremely large, the product will be nearly
worthless. Assume that the unit cost of producing the product is a constant $%
c$ with $c<a$. Then the profits can be written as%
\begin{eqnarray}
u_{j}\left( q_{1},q_{2}\right) &=&q_{j}\left[ P\left( Q\right) -c\right]
\notag \\
&=&q_{j}\left[ a-c-\left( q_{1}+q_{2}\right) \right]  \notag \\
&=&q_{j}\left[ k-\left( q_{1}+q_{2}\right) \right] ,  \label{eq 1}
\end{eqnarray}%
with $k=a-c$ being a constant and $j=1,2$. It is easy to find that the
unique Nash equilibrium of the game is%
\begin{eqnarray}
q_{1}^{\ast }=q_{2}^{\ast }=\frac{k}{3}.
\end{eqnarray}%
At this equilibrium the profits are%
\begin{eqnarray}
u_{1}\left( q_{1}^{\ast },q_{2}^{\ast }\right) =u_{2}\left( q_{1}^{\ast
},q_{2}^{\ast }\right) =\frac{k^{2}}{9}.
\end{eqnarray}%
However this equilibrium fails to be the optimal solution. It is easy to
check that if the two firms can cooperate and restrict their quantities to%
\begin{eqnarray}
q_{1}^{\prime }=q_{2}^{\prime }=\frac{k}{4},
\end{eqnarray}%
they can both acquire higher profits%
\begin{eqnarray}
u_{1}(\frac{k}{4},\frac{k}{4})=u_{2}(\frac{k}{4},\frac{k}{4})=\frac{k^{2}}{8}%
.
\end{eqnarray}%
In fact, this is the highest profits they can ever attained while remaining
the symmetry of the game. But they can never escape the Nash equilibrium
since uniliteral deviation will decrease the individual profit. This
dilemma-like situation again exhibits the conflict between individual
rationality and collective rationality, just like in the classical
Prisoners' Dilemma\cite{2}.

\section{Quantum Cournot's Duopoly}

We now investigate the quantization of the classical Cournot's Duopoly. The
extension of a classical game into the quantum domain is usually done by
setting up a Hilbert space, assigning the possible outcomes of each
classical strategy to certain quantum states. The quantum strategies are
therefore operations on the quantum states, and the payoffs are read out
from the final measurement. Since different classical strategies are
completely distinguishable, it is then natural to require that the Hilbert
space used to have at least the same number of distinguishable states as
that of the different classical strategies. Therefore in the cases that the
strategic space is a continuum, one needs a Hilbert space with a continuous
set of orthogonal bases, \textit{i.e.} the Hilbert space of a
continuous-variable quantum system.

\begin{figure}[t]
\begin{center}
\includegraphics{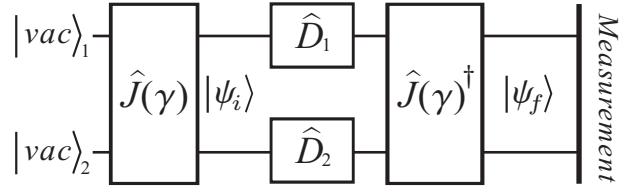}
\end{center}
\caption{The quantum structure of the Cournot's Duopoly.}
\label{Model}
\end{figure}

In this paper, we would like to make use of two single-mode
electromagnetic fields, of which the quadrature amplitudes have a
continuous set of eigenstates. Fig. \ref{Model} shows the quantum
structure of the game. The game starts from state $\left\vert
vac\right\rangle _{1}\otimes \left\vert vac\right\rangle _{2}$,
which is the tensor product of two single-mode vacuum states of
two electromagnetic fields. This state consequently undergoes a
unitary operation $\hat{J}\left( \gamma \right) $, which is known
to both firms (the meaning of $\gamma $ will be presented later).
In order to maintain the symmetry of this game, $\hat{J}\left(
\gamma \right) $ should be symmetric with respect to the
interchange of the two
electromagnetic fields. At this stage, the state of the game is%
\begin{eqnarray}
\left\vert \psi _{i}\right\rangle =\hat{J}\left( \gamma \right) \left\vert
vac\right\rangle _{1}\left\vert vac\right\rangle _{2},
\end{eqnarray}%
Then the two electromagnetic fields are sent to firm $1$ and firm $2$,
respectively. Strategic moves of firm $1$ and firm $2$ are associated with
unitary operators $\hat{D}_{1}$ and $\hat{D}_{2}$, respectively, which are
chosen from certain strategic spaces and are obviously local operators on
their individual electromagnetic fields. Having executed their moves, firm $1
$ and firm $2$ forward their electromagnetic fields to the final
measurement, prior to which a disentangling operation $\hat{J}\left( \gamma
\right) ^{\dagger }$ is carried out. Therefore the final state prior to the
measurement is%
\begin{eqnarray}
\left\vert \psi _{f}\right\rangle =\hat{J}\left( \gamma \right) ^{\dagger }(%
\hat{D}_{1}\otimes \hat{D}_{2})\hat{J}\left( \gamma \right) \left\vert
vac\right\rangle _{1}\left\vert vac\right\rangle _{2}.
\end{eqnarray}%
It is straightforward to set the final measurement be corresponding to the
observables $\hat{X}_{j}=(\hat{a}_{j}^{\dagger }+\hat{a}_{j})/\sqrt{2}$ (the
\textquotedblleft position\textquotedblright\ operators) of firm $j$, where $%
\hat{a}_{j}^{\dagger }$ ($\hat{a}_{j}$) is the creation (annihilation)
operator of firm $j$'s electromagnetic field. This measurement is usually
done by the homodyne measurement. In order to reduce the noise of the result
(since a coherent state is not an eigenstate of the \textquotedblleft
position\textquotedblright\ operator), we shall squeeze the reference light
beam in the homodyne measurement. In this paper we focus on the idea case
that the reference light beam is infinitely squeezed, then the noise is
reduced to zero. Let $\tilde{x}_{j}$ be the measurement result, then the
individual quantity is determined by $q_{j}=\tilde{x}_{j}$, and hence the
profit by%
\begin{eqnarray}
u_{j}^{Q}(\hat{D}_{1},\hat{D}_{2})=u_{j}(\tilde{x}_{1},\tilde{x}_{2}),
\end{eqnarray}%
where superscript \textquotedblleft $Q$\textquotedblright\ denotes
\textquotedblleft quantum\textquotedblright .

The classical Cournot's Duopoly should be included as a subset of the
quantum structure, so that the quantum game is comparable with the original
classical one\cite{2}. Indeed the classical game is faithfully represented
as long as $\hat{J}\left( \gamma \right) =\hat{J}\left( \gamma \right)
^{\dagger }=I$ (the identity operators). Let $\hat{P}_{j}=i(\hat{a}%
_{j}^{\dagger }-\hat{a}_{j})/\sqrt{2}$ be \textquotedblleft
momentum\textquotedblright\ operator of firm $j$'s electromagnetic field. We
can see that if the players are restricted to choose their strategies from
the sets%
\begin{eqnarray}
S_{j}=\{\hat{D}_{j}\left( x_{j}\right) =\exp (-ix_{j}\hat{P}_{j})\left\vert
x_{j}\in \left[ 0,\infty \right) \right. \},\text{ }j=1,2
\end{eqnarray}%
then the final state is%
\begin{eqnarray}
\left\vert \psi _{f}\right\rangle =(\exp (-ix_{1}\hat{P}_{1})\left\vert
vac\right\rangle _{1})\otimes (\exp (-ix_{2}\hat{P}_{2})\left\vert
vac\right\rangle _{2}).
\end{eqnarray}%
And the consequent final measurement gives the results $q_{1}=x_{1}$ and $%
q_{2}=x_{2}$, thus the original classical game is recovered. We hence obtain
that the set $S_{j}$ is exactly the quantum analog of classical strategic
space.

In the remaining part of the paper, we would like to present the
\textquotedblleft minimal\textquotedblright\ extension of the classical
Cournot's Duopoly into the quantum domain: we maintain the strategic space
unexpanded ($S_{j}$ for firm $j$) and only extend the initial state $%
\left\vert \psi _{i}\right\rangle $ to be superposition or entangled state ($%
\hat{J}\left( \gamma \right) \neq I$ and $\hat{J}\left( \gamma \right)
^{\dagger }\neq I$). The classical game is a subset of this
\textquotedblleft minimal\textquotedblright\ extension in the sense that the
quantum game turns back to the original classical form when the initial
state is not entangled. In this extension, if the game of non-zero
entanglement exhibits any features not seen in the classical game, we can be
sure that these features are completely attributed to the quantum
entanglement. However it does not rule out the possibility of getting a
quantum version of Cournot's Duopoly by extending both the states and the
strategic space.

The entangling operator $\hat{J}\left( \gamma \right) $ is given by%
\begin{eqnarray}
\hat{J}\left( \gamma \right) =\exp \{-\gamma (\hat{a}_{1}^{\dagger }\hat{a}%
_{2}^{\dagger }-\hat{a}_{1}\hat{a}_{2})\}=\exp \{i\gamma (\hat{X}_{1}\hat{P}%
_{2}+\hat{X}_{2}\hat{P}_{1})\},
\end{eqnarray}%
Hence the initial state is%
\begin{eqnarray}
\left\vert \psi _{i}\right\rangle =\exp \{-\gamma (\hat{a}_{1}^{\dagger }%
\hat{a}_{2}^{\dagger }-\hat{a}_{1}\hat{a}_{2})\}\left\vert vac\right\rangle
_{1}\left\vert vac\right\rangle _{2},
\end{eqnarray}%
which is exactly the two-mode squeezed vacuum state used to teleport
continuous quantum variables\cite{10,10-1,10-2} and to demonstrate the
violation of Bell's inequalities for continuous variable systems\cite{11}. $%
\gamma \geqslant 0$ is known as the squeezing parameter and can be
reasonably regarded as a measure of entanglement. Note that in the infinite
squeezing limit $\gamma \rightarrow \infty $, the initial state approximates
the EPR state, \textit{i.e.} $\lim_{\gamma \rightarrow \infty }\left\vert
\psi _{i}\right\rangle =\int \left\vert x,-x\right\rangle dx=\left\vert
\text{EPR}\right\rangle $\cite{10,10-1,10-2}.

Detailed calculation gives%
\begin{eqnarray*}
\hat{J}\left( \gamma \right) ^{\dagger }\hat{D}_{1}\left( x_{1}\right) \hat{J%
}\left( \gamma \right) &=&\exp \{-ix_{1}(\hat{P}_{1}\cosh \gamma +\hat{P}%
_{2}\sinh \gamma )\}, \\
\hat{J}\left( \gamma \right) ^{\dagger }\hat{D}_{2}\left( x_{2}\right) \hat{J%
}\left( \gamma \right) &=&\exp \{-ix_{2}(\hat{P}_{2}\cosh \gamma +\hat{P}%
_{1}\sinh \gamma )\}.
\end{eqnarray*}%
Therefore%
\begin{eqnarray*}
&&\hat{J}\left( \gamma \right) ^{\dagger }[\hat{D}_{1}\left( x_{1}\right)
\otimes \hat{D}_{2}\left( x_{2}\right) ]\hat{J}\left( \gamma \right) \\
&=&\exp \{-i(x_{1}\cosh \gamma +x_{2}\sinh \gamma )\hat{P}_{1}\}\exp
\{-i(x_{2}\cosh \gamma +x_{1}\sinh \gamma )\hat{P}_{2}\}.
\end{eqnarray*}%
From this we obtain the final state%
\begin{eqnarray}
\left\vert \psi _{f}\right\rangle &=&\exp \{-i(x_{1}\cosh \gamma +x_{2}\sinh
\gamma )\hat{P}_{1}\}\left\vert vac\right\rangle _{1}  \notag \\
&&\otimes \exp \{-i(x_{2}\cosh \gamma +x_{1}\sinh \gamma )\hat{P}%
_{2}\}\left\vert vac\right\rangle _{2}.  \label{eq 2}
\end{eqnarray}

The final measurement gives the respective quantities of the two firms%
\begin{eqnarray*}
q_{1} &=&x_{1}\cosh \gamma +x_{2}\sinh \gamma , \\
q_{2} &=&x_{2}\cosh \gamma +x_{1}\sinh \gamma .
\end{eqnarray*}%
For convenience, we directly denote the strategy in the quantum game by $%
x_{1}$ and $x_{2}$ when the strategies are $\hat{D}_{1}\left( x_{1}\right) $
and $\hat{D}_{2}\left( x_{2}\right) $, respectively. Referring equation (\ref%
{eq 1}) the quantum profits for them are%
\begin{eqnarray}
u_{1}^{Q}\left( x_{1},x_{2}\right) &=&u_{1}(q_{1},q_{2})  \notag \\
&=&\left( x_{1}\cosh \gamma +x_{2}\sinh \gamma \right) \times \left[
k-e^{\gamma }\left( x_{1}+x_{2}\right) \right] ,  \notag \\
u_{2}^{Q}\left( x_{1},x_{2}\right) &=&u_{2}(q_{1},q_{2})  \notag \\
&=&\left( x_{2}\cosh \gamma +x_{1}\sinh \gamma \right) \times \left[
k-e^{\gamma }\left( x_{1}+x_{2}\right) \right] .  \label{eq 3}
\end{eqnarray}%
Solving the Nash equilibrium gives the unique one%
\begin{eqnarray}
x_{1}^{\ast }=x_{2}^{\ast }=\frac{k\cosh \gamma }{1+2e^{2\gamma }}\text{.}
\label{eq 4}
\end{eqnarray}%
And the profits at this equilibrium are%
\begin{eqnarray}
u_{1}^{Q}\left( x_{1}^{\ast },x_{2}^{\ast }\right) =u_{2}^{Q}\left(
x_{1}^{\ast },x_{2}^{\ast }\right) =\frac{k^{2}e^{\gamma }\cosh \gamma }{%
(3\cosh \gamma +\sinh \gamma )^{2}}\text{.}  \label{eq 5}
\end{eqnarray}

\begin{figure}[tbp]
\begin{center}
\includegraphics{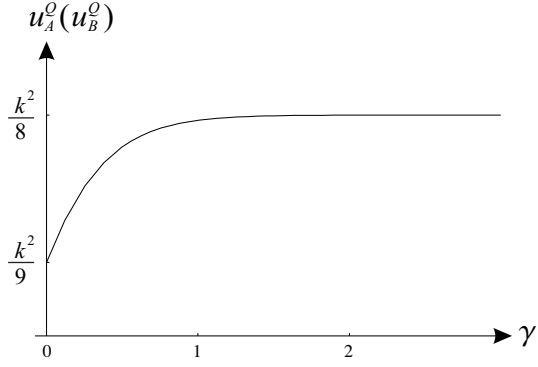}
\end{center}
\caption{The profits at quantum Nash equilibrium as a function of the
squeezing parameter $\protect\gamma $, which can be reasonably regarded as a
measure of entanglement.}
\label{Payoff}
\end{figure}

It would be interesting to review how the profits at Nash equilibrium vary
with respect to the measure of entanglement $\gamma $, as depicted in Fig. %
\ref{Payoff}. It shows that the more entangled the game's state is, the
higher profits the two firms can obtain. When the game is not entangled,
\textit{i.e.} $\gamma =0$, the quantum game goes back to the original
classical form. However in maximally entangled game, \textit{i.e.} in the
infinite squeezing limit $\gamma \rightarrow \infty $, the initial state $%
\left\vert \psi _{i}\right\rangle \rightarrow \left\vert \text{EPR}%
\right\rangle $ and $u_{1}^{Q}\left( x_{1}^{\ast },x_{2}^{\ast }\right)
=u_{2}^{Q}\left( x_{1}^{\ast },x_{2}^{\ast }\right) \rightarrow k^{2}/8$,
which is the best situation the two firms can ever achieve. This novel
feature indicates that the original Einstein-Podolksy-Rosen state enables
the two firms to best cooperate and therefore to be best rewarded, and the
dilemma-like situation in the classical game is completely removed.

The alert reader may argue that the quantum structure proposed here can be
simulated purely classically. For instance, firm $1$ and firm $2$ may each
communicate their choice ($x_{1}$ and $x_{2}$) to the judge using ordinary
telephone lines. The judge then computes the payoffs according to equation (%
\ref{eq 3}). However this implementation loses comparability with the
original game, because once the judge has to change the payoff functions
according to $\gamma $, at the same time he changes the game itself, since
in the original game the payoffs are always given by the same procedure. Yet
the quantum structure maintains the comparability since the payoff are given
by the same procedure independent of $\gamma $: measuring the quadrature
amplitudes, taking the results as quantities, and calculating the profits
according to the classical payoff functions (in equation (\ref{eq 1})).
Further, any classical \textquotedblleft simulation\textquotedblright\
maintaining the comparability should have to introduce direct interaction
between strategies of the players (firms), because the final quantity of a
single firm should contain the strategy of its opponent (see equation (\ref%
{eq 3})). This direct interaction, however, is prohibited since the game is
a \textit{static} one. While in the quantum structure the entanglement
provides an \textquotedblleft invisible channel\textquotedblright\ for the
two firms to affect each other, rather than to directly interact.

\section{Experimental Setup}

\begin{figure}[tbp]
\begin{center}
\includegraphics{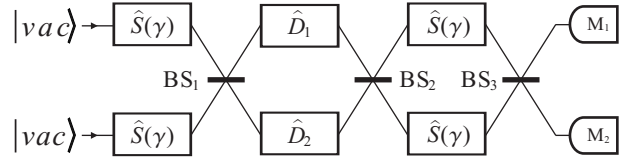}
\end{center}
\caption{Practical setup based on optical experiments for the quantum
structure of Cournot's Duopoly (shown in Fig. \protect\ref{Model}). $\hat{D}%
_{j}$ is the strategic move adopted by firm $j$, and M$_{j}$ denotes the
measurement of firm $j$'s.}
\label{ExpSetup}
\end{figure}

So far all the investigations are in theory, we now want to
construct a practical setup of the quantum Cournot's Duopoly,
based on feasible optical experiments. The setup is shown in Fig.
\ref{ExpSetup}. BS$_{1}$, BS$_{2}$
and BS$_{3}$ are beam splitters. The operator of a beam splitter is\cite%
{12,12-1}%
\begin{eqnarray}
\hat{B}(\theta ,\phi )=\exp \{\frac{\theta }{2}(\hat{a}_{1}^{\dag }\hat{a}%
_{2}e^{i\phi }-\hat{a}_{1}\hat{a}_{2}^{\dag }e^{-i\phi })\}\text{,}
\end{eqnarray}%
with the amplitude reflection and transmission coefficients $t=\cos \theta
/2 $ and $r=\sin \theta /2$, and $\phi $ being the phase difference between
the reflected and transmitted fields. Here we set%
\begin{eqnarray*}
\text{BS}_{1} &=&\text{BS}_{2}=\hat{B}\left( \pi /2,3\pi /2\right) , \\
\text{BS}_{3} &=&\hat{B}\left( \pi /2,\pi /2\right) =\text{BS}_{2}^{\dag }.
\end{eqnarray*}%
$\hat{S}\left( \gamma \right) =\exp \{-i\gamma (\hat{a}^{\dag 2}+\hat{a}%
^{2})/2\}$ is squeezing operator, which can be implemented by parametric
down-conversion inside a nonlinear crystal\cite{13}. Detailed calculation
yields that%
\begin{eqnarray}
\text{BS}_{1}e^{-i\frac{\gamma }{2}(\hat{a}_{1}^{\dag 2}+\hat{a}%
_{1}^{2})}e^{-i\frac{\gamma }{2}(\hat{a}_{2}^{\dag 2}+\hat{a}%
_{2}^{2})}\left\vert vac \right\rangle \left\vert vac \right\rangle &=&e^{-\gamma (\hat{a}_{1}^{\dagger }%
\hat{a}_{2}^{\dagger }-\hat{a}_{1}\hat{a}_{2})}\left\vert vac
\right\rangle \left\vert vac \right\rangle
\text{,}  \notag \\
\text{BS}_{3}e^{-i\frac{\gamma }{2}(\hat{a}_{1}^{\dag 2}+\hat{a}%
_{1}^{2})}e^{-i\frac{\gamma }{2}(\hat{a}_{2}^{\dag 2}+\hat{a}_{2}^{2})}\text{%
BS}_{2} &=&e^{\gamma (\hat{a}_{1}^{\dagger }\hat{a}_{2}^{\dagger }-\hat{a}%
_{1}\hat{a}_{2})}\text{.}
\end{eqnarray}%
Therefore the operations of $\hat{J}\left( \gamma \right) $ and $\hat{J}%
\left( \gamma \right) ^{\dag }$ are constructed.$\ \hat{D}_{1}(x_{1})$ and $%
\hat{D}_{2}(x_{2})$ can be realized by simple phase-space
displacements on each single electromagnetic field. The setup in
Fig. \ref{ExpSetup} hence faithfully represents the quantum
structure in Fig. \ref{Model}.

\section{Conclusion}

We investigate the quantization of games with continuum strategic
space, making use of continuous-variable quantum systems. For the
particular case of Cournot's Duopoly, we construct a quantum
structure of it, and proposed the extension from classical to
quantum, which we refer to be \textquotedblleft
minimal\textquotedblright . The classical game is shown to be a
subset of the quantum game, and therefore they can be compared in
an unbiased manner. We observed novel features in the quantum
Cournot's Duopoly, which are completely due to quantum
entanglement. In the quantum game two firms virtually cooperate
even though they still behave selfishly. This virtual cooperation,
as well as the profits, increases as the entanglement increases.
For the maximal entanglement limit, the initial state is precisely
the original Einstein-Podolksy-Rosen state, and enables the two
firms to cooperate best and to be best rewarded. Further we
proposed a feasible experimental scheme to demonstrate the quantum
Cournot's Duopoly, showing the game's procedure of preparing
initial states, executing strategic moves and reading out payoffs
(profits). This scheme is within the capability of current optical
technology, and could be implemented indeed.

We thank Z. B. Chen for helpful discussion. This work was
supported by the National Nature Science Foundation of China
(Grant No. 10075041), the National Fundamental Research Program
(2001CB309300), and the ASTAR Grant No. 012-104-0040.

\end{document}